\documentclass[a4paper,11pt]{article}
\usepackage{pos}

\newcommand{\stripper}{\textsc{STRIPPER}}
\newcommand{\mulbos}{\textsc{MulBos}}
\newcommand{\powhegbox}{\textsc{Powheg-Box-Res}}
\newcommand{\pythia}{\textsc{Pythia}}
\newcommand{\sherpa}{\textsc{Sherpa}}
\newcommand{\madgraph}{\textsc{MG5\_aMC}}

\newcommand{\bib}[1]{Ref.~\cite{#1}}

\newcommand{\fig}[1]{Fig.~\ref{#1}}

\newcommand{\sect}[1]{Section~\ref{#1}}

\newcommand{\gev}{{\unskip\,\text{GeV}}}

\title{Theory overview on polarization and spin-correlations in multi-boson processes}

\author[a]{Thi Nhung Dao}
\author*[a]{Duc Ninh Le}

\affiliation[a]{Phenikaa Institute for Advanced Study, Phenikaa University, \\
  Hanoi 12116, Vietnam}

\emailAdd{nhung.daothi@phenikaa-uni.edu.vn}
\emailAdd{ninh.leduc@phenikaa-uni.edu.vn}

\abstract{
In this contribution, we summarize and discuss recent theoretical progress 
on the Standard-Model calculation of joint polarized cross sections of massive multi-boson processes 
with fully leptonic decays. 
The topics include fixed-order calculation of 
higher order QCD and electroweak corrections for both production 
and decay amplitudes, and parton-shower effects.   
}

\FullConference{The Thirteenth Annual Large Hadron Collider Physics (LHCP2025)\\
5-9 May 2025\\
Taipei, Taiwan\\}

\begin{document}
\maketitle

\section{Introduction}
Multi-boson production processes at the Large Hadron Collider (LHC) 
can play an important role 
in the quest 
for a better understanding of  
the Electroweak (EW) symmetry breaking mechanism 
and physics beyond the Standard Model (SM). 
This is because the purely leptonic final states offer feasible paths amid a challenging hadronic environment. 
Moreover, semi-leptonic final states and their interplay with the fully leptonic final states can be 
of great help in these studies.  

A steady progress of measurements on diboson production has been going on 
from LEP \cite{ALEPH:2013dgf}, Tevatron \cite{D0:2009xgd,CDF:2016zte} to the LHC. The diboson measurement accuracy, 
for integrated unpolarized fiducial cross sections 
at around $140$ $\text{fb}^{-1}$ of integrated luminosity of Run-2, is about $4\%$ for both ATLAS $WZ$ \cite{ATLAS:2025edf}, $W^+W^-$ \cite{ATLAS:2025dhf} and CMS $ZZ$ \cite{CMS:2020gtj}, $WZ$ \cite{CMS:2021icx}. 
First observations and measurements of triboson processes have also been 
coming, see ATLAS $VVZ$, $W^+W^-Z$ \cite{ATLAS:2024nab} and CMS $W^+W^-Z$ \cite{CMS:2025hlu}. 
The uncertainties of these measurements are still quite large due to the smallness of the 
cross sections. They will however be reduced in the future with more data, better measurement and analysis techniques. 

In parallel with the unpolarized cross section measurements, there have been also attempts to measure joint polarized 
cross sections of massive diboson production processes from LEP via $W^+W^-$ \cite{OPAL:2000wbs,L3:2003tnr,DELPHI:2009wdg} to the LHC 
via same-sign $WWjj$ \cite{CMS:2020etf,ATLAS:2025wuw}, $ZZ$ \cite{ATLAS:2023zrv}, and $WZ$ \cite{ATLAS:2024qbd}. 
Single polarization of a massive gauge boson has also been measured, see e.g. $WZ$ \cite{Aaboud:2019gxl,CMS:2021icx}. 
These measurements are important and can serve as a bridge between the joint polarization and 
unpolarization. Since the focus of this contribution is on joint polarization, the case 
of single polarization will not be discussed further here. 

Together with these experimental efforts, precise theoretical predictions and precise simulation tools are needed. 
The most precise theoretical predictions are obtained via fixed-order calculations. 
These calculations are however very challenging and are limited to a few number of particles 
in the final state. For the joint polarization case,
the state of the art is at next-to-next-to-leading order (NNLO) QCD level for diboson 
and at next-to-leading order (NLO) QCD+EW for vector boson scattering \cite{Denner:2024tlu}. 

In order to compare with real data, where the number of final-state particles is uncountable, additional steps must be appended 
to the fixed-order calculation. These include QCD and QED showers, hadronization, and detector simulation.  
In this contribution, we will summarize latest theoretical results for the fixed-order calculations and parton showers (PS).  

\section{NNLO QCD effects}
\label{nnlo_qcd}
For inclusive polarized diboson production,  
NNLO QCD corrections were calculated 
in \cite{Poncelet:2021jmj} for the case of $W^+W^-$, and first results 
for $ZZ$ have been recently presented in \cite{Carrivale:2025mjy}. 
Similar results for the $WZ$ case are still missing.

These results are based on two main ingredients. 
The first one is the NNLO QCD calculation techniques developed for 
the unpolarized calculation, in particular the 
novel subtraction scheme for double-real radiation at NNLO \cite{Czakon:2010td,Czakon:2014oma} 
and the two-loop $q\bar{q}$ amplitudes \cite{Gehrmann:2015ora}.
The second ingredient is the double-pole approximation (DPA) \cite{Aeppli:1993cb,Aeppli:1993rs,Denner:2000bj,Denner:2021csi} needed 
to separate the polarizations.

\begin{figure}[t!]
  \centering
  \includegraphics[width=0.49\textwidth]{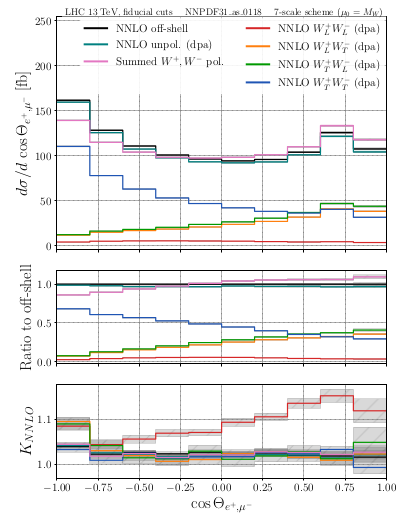}
  \includegraphics[width=0.49\textwidth]{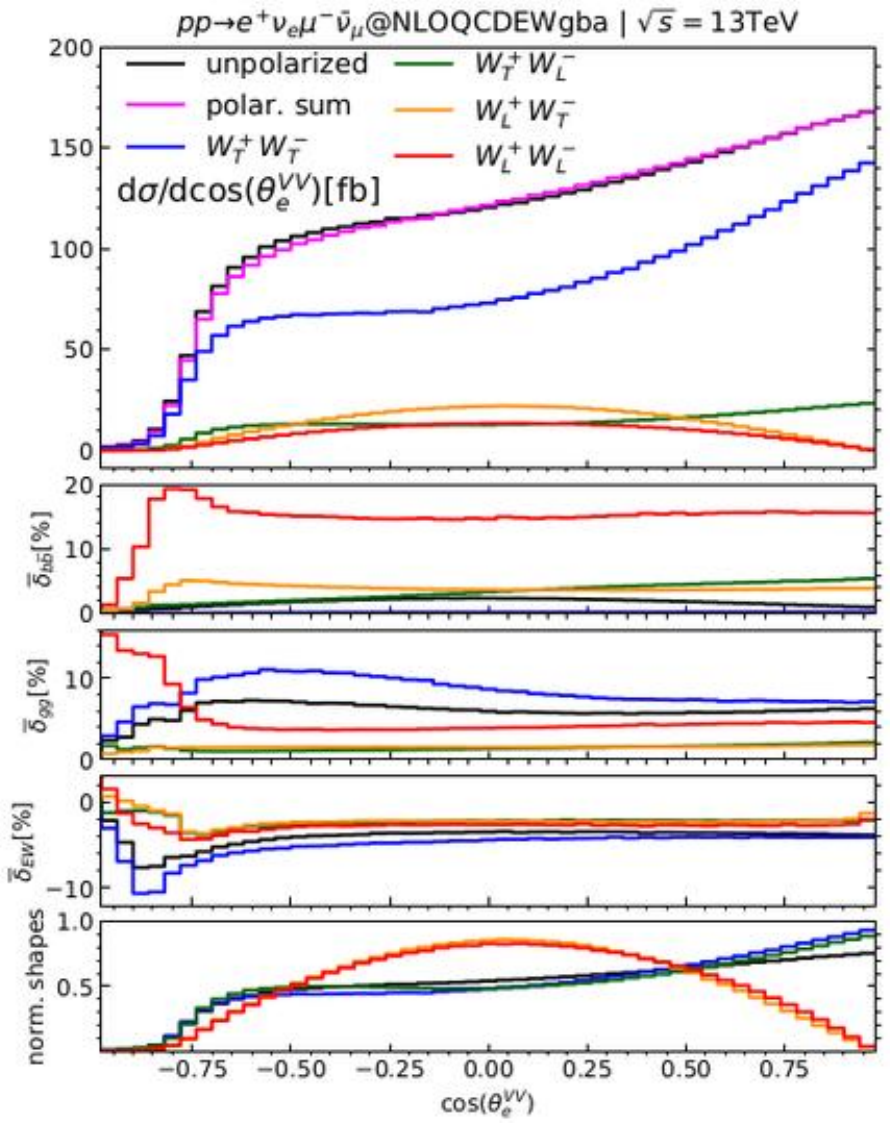}\\
  \includegraphics[width=0.50\textwidth]{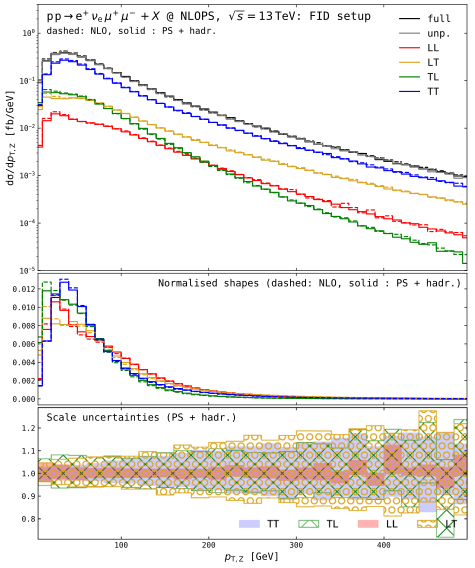}
  \caption{Top left: Distributions in $\cos\theta_{e^+,\mu^-}$ (the angle between the two leptons) at NNLO QCD for the inclusive $W^+W^-$ production, produced by \stripper\ \cite{Poncelet:2021jmj}. Top right: Distributions in $\cos(\theta_e^{VV})$ (see text) at NLO QCD+EW also for the inclusive $W^+W^-$ production, produced by \mulbos\ \cite{Dao:2023kwc}. 
  Bottom: Distributions in $p_{T,Z}$ at NLO QCD merged with PS for the inclusive $W^+Z$ production, 
  produced by \powhegbox\ \cite{Pelliccioli:2023zpd}.}
  \label{fig:dis_nnloQCD_nloEW_PS}
\end{figure}
These ingredients have been combined in the Monte-Carlo program \stripper\ \cite{Poncelet:2021jmj,Carrivale:2025mjy}, 
which has been used to produce the NNLO QCD results for $W^+W^-$ and $ZZ$. \stripper\ can produce differential cross sections, such as the one shown in \fig{fig:dis_nnloQCD_nloEW_PS} (top left). 
\section{NLO EW effects}  
\label{nlo_qcd_ew}
For polarized cross sections, NLO EW corrections are complicated due to photon emission off an intermediate 
$W^\pm$ boson which can be on-shell (OS). The bottleneck is to define polarizations of real-emission amplitudes which are soft divergent. 
The collinear divergences do not occur here because the OS $W$ bosons are massive.  

The state of the art calculation is the same-sign $W^+W^+jj$ process \cite{Denner:2024tlu}. 
We will however take the case of inclusive $WZ$ production as an example for the sake of simplicity. 
Polarizations are defined using the DPA amplitudes, which are 
functions of propagator factors, production amplitudes, and decay amplitudes. 
The propagator factors come from the intermediate gauge bosons ($W$ and $Z$ in this case). 
They depend on the off-shell momentum and the total decay width of the gauge boson. 
The production and decay amplitudes must be OS quantities as required by gauge invariance. 
This OS requirement is achieved as follows. 
First, generate {\em all} Feynman diagrams for the OS production processes $q^\prime\bar{q} \to WZ$ and 
$q^\prime\bar{q} \to WZ \gamma$, and OS decay processes $V_i \to ll$ and $V_i \to ll \gamma$. 
The expressions for these amplitudes are then obtained as functions of the external momenta and helicities. 
These momenta must satisfy the OS conditions, i.e. $\hat{p}_i^2 = m_i^2$, where the hat 
is used to indicate the on-shellness. 
Second, the OS momenta $\hat{p}_i$ are calculated from the off-shell momenta $p_i$ using OS mappings. 
These mappings are not unique, but the differences are of the order of $\mathcal{O}(\Gamma_V/M_V)$ 
being the intrinsic uncertainty of the DPA. 

Those ingredients are needed to build the DPA amplitudes of the photon-emission 
$q^\prime\bar{q} \to e^+\nu_e \mu^+ \mu^- \gamma$ process. 
These amplitudes contain soft divergences in the limit $p^\mu_\gamma \to 0$. 
These divergences cancel with the corresponding ones in the virtual corrections. 
In practice, we introduce intermediate IR-subtraction terms to make the 
virtual-correction part and the real-emission part separately finite. 
The Catani-Seymour (CS) subtraction method has been used for this purpose in all 
relevant publications $WZ$ \cite{Le:2022lrp,Le:2022ppa}, $WW$ \cite{Denner:2023ehn,Dao:2023kwc}, and same-sign $W^+W^+jj$ \cite{Denner:2024tlu}. 
The integrated subtraction term of the virtual-correction part is easy as the photon momentum vanishes here and 
the kinematics is Born-like. 
However, constructing the DPA amplitudes for the subtraction term 
of the real-emission part is a bottleneck. 
For this local subtraction term, we need OS momenta for the singular factor (denoted as $\hat{g}_\text{sub}$ in Eq. (4.13) of 
\cite{Le:2022ppa}, while as $\mathcal{D}_{ij,k}$ in Eq. (B.5) of \cite{Denner:2024tlu})
for the reduced DPA amplitude $\mathcal{B}_{e^+\nu_e \mu^- \bar{\nu}_\mu}$ (notice the absence of the photon). 
The corresponding off-shell momenta are also needed for the propagator factors in 
the $\mathcal{B}$ and in the kinematic cuts. 
For the reduced DPA Born amplitude, one first applies the CS mapping to the original 
off-shell momenta $p_i$ to obtain $\tilde{p}_i$ (the tilde means this is a CS-mapped momentum), and 
then the OS mapping to obtain $\hat{\tilde{p}}_i$. This is done in both groups \cite{Le:2022ppa} and \cite{Denner:2024tlu}. 
For the the singular factor, \bib{Le:2022ppa} uses the OS momenta $\hat{p}_i$ 
calculated directly from the off-shell $p_i$, 
while \bib{Denner:2024tlu} uses $\hat{\tilde{p}}_i$ (i.e. OS mapping after CS mapping). 
These different procedures may lead to different results. 
The differences are however expected to be very small. Detailed numerical investigations are still 
needed to quantify this. 

There is another difference between \bib{Le:2022ppa} and \bib{Denner:2024tlu} in the 
application of kinematic cuts for the local dipole subtraction term. 
The CS momenta $\tilde{p}_i$ (obtained by applying the CS mapping on the off-shell $p_i$) are used 
in \cite{Denner:2024tlu}, while $\tilde{p}^\prime_i$ are used in \cite{Le:2022ppa}. 
The leptonic $\tilde{p}^\prime_i$ are calculated from the off-shell $\tilde{p}_W$ by 
letting it decays into two massless particles. 
This decaying procedure is detailed in \cite{Le:2022ppa} (see Eq. (4.17) and the paragraph after it). 
Since the $\tilde{p}^\prime_i$ are different from the $\tilde{p}_i$, 
a second source of differences comes from the mismatch related to these leptonic momenta occurring 
in the local subtraction term (such as the kinematic cuts or dynamical scales). 
These differences are again expected to be very small, being of the order of the intrinsic 
uncertainty of the DPA. 

A comparison between these two approaches was done in \cite{Dao:2024jon} (see Table 3 there) for the case of $W^+W^-$ production. 
The differences for the integrated NLO EW polarized cross sections are $0.01\%$ for the LL, $0.1\%$ for the mixed polarization 
(LT or TL), $0.4\%$ for the TT. These differences are within the DPA uncertainty. 

To illustrate NLO EW effects, we show in \fig{fig:dis_nnloQCD_nloEW_PS} (top right) the polarized distributions in $\cos\theta_e^{VV}$ of 
the $W^+W^-$ production, 
where $\theta_e^{VV}$ is the angle between $\vec{p}_e$ (determined in the $W^+$ rest frame) 
and $\vec{p}_{W^+}$ (determined in the $W^+W^-$ center of mass frame). This distribution 
is important for the polarization separation. More details are provided in \cite{Dao:2023kwc}. 
\section{Parton shower effects}
\label{sect_ps}
PS comes after the hard process and before hadronization. 
It introduces QCD (light quarks and gluon) radiation and QED (photon) radiation. 
These parton showers are based on soft and collinear approximations (see this recent review \cite{Campbell:2022qmc} 
for details).

For the case of polarized multi-boson production with fully leptonic decays, 
QCD shower comes solely from initial state radiation (ISR), 
while QED shower comes from ISR, final state radiation (FSR), 
and intermediate state radiation if the process involves a $W$ boson (see \sect{nlo_qcd_ew}). 
The QED effects are expected to be much smaller than the QCD ones at the LHC due 
to the smallness of the QED coupling strength. 

Since the OS mapping needed to separate polarized cross sections affects only 
the momenta of the decay products (leptons and photons) and leaves 
the momenta of the colored particles untouched, the QCD shower is the same 
for all polarized cross sections as well as the unpolarized one. 
One can therefore use the existing PS implementation for the 
unpolarized case. The changes needed are replacing the unpolarized hard amplitudes 
by a sum of polarized amplitudes using the DPA, and implementing the OS mapping. 
These changes are the same for the fixed order calculations and have 
been worked out in detail in \cite{Denner:2021csi}. 
QCD PS matched with NLO QCD corrections for all diboson processes, using the 
\powhegbox\ framework \cite{Nason:2004rx,Frixione:2007vw,Alioli:2010xd,Jezo:2015aia} 
has been achieved in \cite{Pelliccioli:2023zpd}. 
\powhegbox\ is a framework to match a NLO calculation with a PS. 
In this case, the PS is done by using \pythia 8 \cite{Sjostrand:2014zea,Sjostrand:2019zhc}.

While implementing the QCD PS is fairly straightforward, the QED PS is more involved 
for polarized processes due to the requirement that the intermediate $W$ bosons must be 
on-shell, hence soft photon radiation from these $W$ bosons must be resummed. 
Since the DPA amplitudes are factorized into a production amplitude and decay amplitudes, 
QED PS must be done separately for the production part and the decays. 
This would give the correct photon-radiation correlations between the ISR and 
the radiation off the $W$ bosons for the production part, and 
correct correlations between the FSR and the photon radiation off the decaying particle, 
as required by gauge invariance. 
This would require a non-trivial change in the interface between the hard amplitudes and the PS tool. 

Numerically, \bib{Pelliccioli:2023zpd} found that, 
for the  $p_{T,Z}$ distribution of the $W^+Z$ production with 
the fiducial cut setup, NLO QCD cross sections matched to QCD PS can deviate from the 
fixed order NLO ones by $10\%$ at $p_{T,Z}\approx 200\gev$.  
An illustration of NLOPS result is shown in \fig{fig:dis_nnloQCD_nloEW_PS} (bottom), 
including both QCD and QED PS.

There is another powerful tool which can provide PS effects for polarized multiboson 
production, namely \sherpa\ \cite{Hoppe:2023uux} (see \cite{Carrivale:2025mjy} for a short summary). 
\sherpa\ can provide PS-matched polarised predictions up to approximate NLO QCD. 
QED shower is also available. 
The polarized events are however generated using 
a spin-correlated narrow-width approximation (NWA) \cite{Hoppe:2023uux}.

\madgraph\ (MadGraph 5 a Monte Carlo at NLO), heavily used by 
ATLAS and CMS collaborations, can also generate polarized diboson samples at LO \cite{Alwall:2014hca,BuarqueFranzosi:2019boy}
(see \cite{Carrivale:2025mjy} for a short summary).
PS effects can be included via an interface to \pythia\ \cite{Sjostrand:2019zhc} using 
the MLM scheme \cite{Mangano:2006rw,Alwall:2007fs}. 
The polarization is separated using either a diagram selection method or a spin-correlated NWA \cite{Carrivale:2025mjy}. 
The first comparison between these codes has recently been done for the $ZZ$ case in \cite{Carrivale:2025mjy}.
\section{Conclusions}
\label{sect_con}
We have discussed and highlighted some recent theoretical results for joint polarized cross sections of multi-boson 
production processes. Particular attentions have been paid to the NNLO QCD, NLO EW, and parton-shower effects.   
Though the discussions were mostly focused on the inclusive diboson case, the ideas can also 
be applied for other classes of processes such as the triboson and vector boson scattering. 
Though impressive results have been achieved, further works are still needed to combine all 
those effects into automated simulation tools, thereby facilitating the simulation process and 
obtaining more precise measurements.     

\acknowledgments
DNL would like to thank the session conveners for their kind invitation 
and the conference organizers for financial support. 
This research is funded by Phenikaa University under grant number PU2023-1-A-18.



\providecommand{\href}[2]{#2}\begingroup\raggedright\endgroup
\end{document}